\documentclass{article}%
\usepackage{amssymb}
\usepackage{amsmath}
\usepackage{graphicx}
\usepackage{amsfonts}%
\providecommand{\U}[1]{\protect\rule{.1in}{.1in}}
\begin{document}
\begin{center}
\textbf{Kinetic models of a K channel with 2 stage voltage sensor activation
and an opening step}

S. R. Vaccaro

\end{center}

\emph{E-mail address:}\newline\emph{srvaccaroadelaide@gmail.com}\newline
%\vspace{0.2in}

\begin{quotation}
The activation of a  K channel with 2 voltage sensors and a separate opening
step may be described by a 4 state kinetic equation. If $k_{o}$ and
$k_{c}$ are the opening and closing rates, and $\alpha$ and $\beta$ are the
voltage sensor rates such that $k_{o}$ $\gg\alpha,k_{c}\gg\beta$  and
$k_{o}$ $\gg k_{c}$, the kinetic equation may be reduced to a 3
state model where the effective backward rate from the activated-open state
is $2\beta k_{c}/(k_{o}+k_{c})$. However, if a K channel voltage
sensor activates in 2 stages, and there is a separate opening step, the
occupation probabilities of the states of the channel satisfy a 10 state
kinetic equation. When the time constants of the characteristic equation for
each voltage sensor differ by an order of magnitude, and $k_{o}$ $\gg k_{c}$,
the open state probability is proportional to $n^{2}$, where
$n$ satisfies a rate equation, and the derived rates have a similar voltage
dependence to the rates $\alpha_{n}$ and $\beta_{n}$ from voltage clamp data.
\end{quotation}

\bigskip

\textbf{INTRODUCTION}

Although early models of the time dependence of the activation of K current
were described by $n^{4}$ where $n$ \ satisfies a rate equation \cite{hh},
subsequent studies \ have shown that there is a separate opening step
following the activation of the S4 sensors, \ based on the motion of gating
charge \ before the opening of the channel \cite{wb}. A separate opening step
has been supported by single channel recordings\ of K channels where there is
a rapid flickering between open and closed states \cite{cn}, and by a Sh5
mutation of Shaker A-type K channels that modifies the voltage dependence of
activation but has no effect on the open-time distribution \ or the transition
rates into and out of the open state \cite{za}.

For a single transition of the K channel voltage sensor to the open state, the
time-dependence of the gating current may be described by an exponential, and
therefore, a rising phase in the gating current was an early indicator of \ a
\ two stage process in the activation of the voltage sensor \cite{wb,bwt}. For
a Na channel, in order to account for several kinetic components of the gating
current, it was also proposed that the S4 sensor in each domain activates in
two steps \cite{rdk}. The gating mechanisms of two stage activation of each
voltage sensor and a separate opening step were incorporated into a kinetic
model of the Shaker IR K channel that was constrained by the measurement of
ionic current, gating current and single channel current \cite{zha}.

A voltage sensor of an ion channel is thermally activated across barriers
within a gating pore that are dependent on the membrane electric field, a
dielectric boundary force and electrostatic forces between positively charged
residues on the S4 segment \ and negatively charged residues on the S1, S2 and
S3 segments \cite{llg}. The solution of the Smoluchowski equation for the
probability density of the voltage sensor in an energy landscape with two
large barriers is a kinetic equation that describes the time dependence of the
occupation probabilities $P_{1\text{ \ }}$to $P_{3\text{ \ }}$\ of each state
\cite{vac1,vac2}. When the time constants of the characteristic equation
differ by an order of magnitude, the kinetic equation reduces to a rate
equation with an effective forward rate function that has a voltage dependence
that is exponential-quasilinear when the equivalent charge of the
rate-limiting transition is small \cite{vac3}.

In this paper, we consider the effect of a separate opening step\ and two
stage activation on the effective rate functions $\alpha_{e}$ and $\beta_{e}$
for a K channel with two voltage sensors. If each voltage sensor activates in
2 stages with forward rates $\alpha\ll\gamma$, and backward rates$\ \ \beta
\gg\delta$, the effective voltage sensor rate function from the activated-open
state \ to the first closed state is $2\beta_{r}k_{o\text{ }}/(k_{c}%
+k_{o\text{ }})$ where $k_{o\text{ }}$ $\gg k_{c}$ are the opening and closing
rates, and $\alpha_{r}$ and \ $\beta_{r}$\ are the reduced rates. The \ rate
functions $\alpha_{e}$ and $\beta_{e}$ \ are determined by equating the
activated-open probability of the kinetic equation and the open probability
$n^{2}$\ \ of a Hodgkin-Huxley (HH) three state model where $n$\ \ satisfies a
rate equation. For each membrane potential, $\alpha_{e}\approx\alpha_{r}$ for
activation when $\alpha_{e}$ $>$\ $\beta_{e}$, and $\ \beta_{e}\ \approx
\beta_{r}k_{c}/(k_{o\text{ }}+k_{c})$ for deactivation when $\alpha_{e}$
$\leq$\ $\beta_{e}$, and therefore, the variation of $\beta_{e}$ relative to
$\alpha_{e}$ is small, and the derived rates have a similar voltage dependence
to the rates $\alpha_{n}$ and $\beta_{n}$ from voltage clamp data \cite{hh}.

\textbf{A KINETIC MODEL WITH TWO VOLTAGE SENSORS AND AN OPENING STEP}

\bigskip

The HH model assumes that the K channel attains an open or conducting state
when the last of the four gating transitions is activated. Later models
\ proposed that a separate opening step is required because the earlier
activation steps are voltage dependent but the transition into the open state
is rapid and weakly voltage dependent. For an ion channel with two identical independent
voltage sensors and a separate opening step, with transition rates $k_{o\text{
}}$ and $k_{c}$, \ the occupation probability of the open state \ $O_{4}$, and
the closed states \ $C_{1}$ to \ $C_{3}$\ \ are determined by the kinetic
equation (see Fig. 1)%

\begin{align}
\frac{dC_{1}}{dt}  &  =-2\alpha C_{1}(t)+\beta C_{2}(t)\label{open1}\\
\frac{dC_{2}}{dt}  &  =2\alpha C_{1}(t)+2\beta C_{3}(t)-(\alpha+\beta
)C_{2}(t)\label{open2}\\
\frac{dC_{3}}{dt}  &  =\alpha C_{2}(t)+k_{c}O_{4}(t)-(2\beta+k_{o\text{ }%
})C_{3}(t)\label{open3}\\
\frac{dO_{4}}{dt}  &  =k_{o\text{ }}C_{3}(t)-k_{c}O_{4}(t). \label{open4}%
\end{align}
where \ $\alpha$ $\ $and $\ \beta$\ are the transition rates of the voltage
sensor. If \ $k_{o\text{ }}$ $\gg k_{c}$,\ $k_{o\text{ }}\gg\alpha$\ \ and
$k_{c\text{ }}\gg\beta$ \ \ the kinetic model may be reduced to (see Fig. 2)%

\begin{align}
\frac{dC_{1}}{dt}  &  =-2\alpha C_{1}(t)+\beta C_{2}(t)\label{3 state1}\\
\frac{dC_{2}}{dt}  &  =2\alpha C_{1}(t)+\frac{2\beta k_{c}}{k_{o\text{ }%
}+k_{c}}\overline{O}_{3}(t)-(\alpha+\beta)C_{2}(t)\label{3 state2}\\
\frac{d\overline{O}_{3}}{dt}  &  =\alpha C_{2}(t)-\frac{2\beta k_{c}%
}{k_{o\text{ }}+k_{c}}\overline{O}_{3}(t), \label{3 state3}%
\end{align}
where $\overline{O}_{3}=C_{3}+O_{4}$.

For K channel activation during a clamp at membrane voltage\ $V$, \ from the
solution of Eqs. (\ref{3 state1}) to (\ref{3 state3}) and $\overline{O}%
_{3}(0)=0$
\begin{equation}
\overline{O}_{3}(t)=\frac{2\alpha^{2}}{\omega_{1}\omega_{2}}\left(
1-\frac{\omega_{2}}{\omega_{2}-\omega_{1}}\exp(-\omega_{1}t)+\frac{\omega_{1}%
}{\omega_{2}-\omega_{1}}\exp(-\omega_{2}t)\right)  , \label{openred}%
\end{equation}
where $\omega_{1}$ and $\omega_{2}$ satisfy the chacteristic equation%

\begin{equation}
\omega^{2}-\omega\left(  3\alpha+\beta+\frac{2\beta k_{c}}{k_{o\text{ }}%
+k_{c}}\right)  +2\alpha^{2}+\frac{2\beta k_{c}}{k_{o\text{ }}+k_{c}}\left(
2\alpha+\beta\right)  =0. \label{char1}%
\end{equation}
The activated-open state probability $\overline{O}_{3}(t)$ during activation may be approximated by%

\begin{equation}
O_{3H}(t)=\left(  \frac{\alpha_{e}}{\alpha_{e}+\beta_{e}}\right)  ^{2}%
(1-\exp(-\left(  \alpha_{e}+\beta_{e}\right)  t))^{2} \label{open3h}%
\end{equation}
where $\ $ $O_{3H}(t)$ is a solution of the kinetic equation%

\begin{align}
\frac{dC_{1}}{dt} &  =-2\alpha_{e}C_{1}(t)+\beta_{e}C_{2}%
(t)\label{3 state hh1}\\
\frac{dC_{2}}{dt} &  =2\alpha_{e}C_{1}(t)+2\beta_{e}O_{3H}(t)-(\alpha
_{e}+\beta_{e})C_{2}(t)\label{3 state hh2}\\
\frac{dO_{3H}}{dt} &  =\alpha_{e}C_{2}(t)-2\beta_{e}O_{3H}%
(t),\label{3 state hh3}%
\end{align}
and the rates \ \ $\alpha_{e}$ $>$\ $\beta_{e}$ for activation may be
determined from the equations%
\begin{align}
\left(  \frac{\alpha_{e}}{\alpha_{e}+\beta_{e}}\right)  ^{2} &  =\frac
{2\alpha^{2}}{\omega_{1}\omega_{2}}\label{o3hs}\\
(1-\exp(-\left(  \alpha_{e}+\beta_{e}\right)  t)^{2} &  =1-\frac{\omega_{2}%
}{\omega_{2}-\omega_{1}}\exp(-\omega_{1}t)+\frac{\omega_{1}}{\omega_{2}%
-\omega_{1}}\exp(-\omega_{2}t).\label{o3ht}%
\end{align}
If  $\beta\approx0$, from Eqs. \ (\ref{char1}), (\ref{o3hs}) and (\ref{o3ht}),
$\omega_{1}\approx\alpha,$\ \ $\omega_{2}\approx2\alpha$, $\alpha_{e}%
\approx\alpha,$ and $\beta_{e}\approx0$.

\ 

From the solution of Equations (\ref{3 state1}) to (\ref{3 state3}) for
deactivation and $\overline{O}_{3}(0)=1$
\[
\overline{O}_{3}(t)=\frac{2\alpha^{2}}{\omega_{1}\omega_{2}}+\frac{1}%
{\omega_{1}(\omega_{2}-\omega_{1})}\left(  \frac{2\beta k_{c}}{k_{o\text{ }%
}+k_{c}}\right)  \left(  \omega_{2}-\alpha-\frac{2\beta k_{c}}{k_{o\text{ }%
}+k_{c}}\right)  \exp(-\omega_{1}t)
\]

\begin{equation}
-\frac{1}{\omega_{2}(\omega_{2}-\omega_{1})}\left(  \frac{2\beta k_{c}%
}{k_{o\text{ }}+k_{c}}\right)  \left(  \omega_{1}-\alpha-\frac{2\beta k_{c}%
}{k_{o\text{ }}+k_{c}}\right)  \exp(-\omega_{2}t). \label{deact sol2}%
\end{equation}
where \ $\overline{O}_{3s}=2\alpha^{2}/(\omega_{1}\omega_{2})$, and
\ $\omega_{1}$, $\omega_{2}$ satisfy the characteristic equation
(\ref{char1}). \ As $\omega_{2}>\omega_{1}$, and $\omega_{1}\approx
\alpha+2\beta k_{c}/(k_{o\text{ }}+k_{c})$, \ %

\begin{equation}
\overline{O}_{3}(t)\approx\frac{2\alpha^{2}}{\omega_{1}\omega_{2}}+\frac
{1}{(\omega_{2}-\omega_{1})}\left(  \omega_{2}-\alpha-\frac{2\beta k_{c}%
}{k_{o\text{ }}+k_{c}}\right)  \exp(-\omega_{1}t)\text{.} \label{deactapprox}%
\end{equation}
However, the solution $O_{3H}(t)$\ \ of Eqs. (\ref{3 state hh1}) to
(\ref{3 state hh3}) for deactivation is%

\begin{equation}
O_{3H}(t)=\frac{(\alpha_{e}+\beta_{e}\exp(-\left(  \alpha_{e}+\beta
_{e}\right)  t))^{2}}{\left(  \alpha_{e}+\beta_{e}\right)  ^{2}},
\label{open3hd}%
\end{equation}
\ and from Eq. (\ref{deact sol2}), if the steady state open probability
\ $\overline{O}_{3s}=2\alpha^{2}/\omega_{1}\omega_{2}\approx0$ and $\alpha
_{e}$ $\leq$\ $\beta_{e}$,\ \ \ during a large hyperpolarisation,
$\ \overline{O}_{3}(t)\approx O_{3H}(t)$ when the rates \ \ $\alpha_{e}$ and
\ $\beta_{e}$ satisfy the equations%

\begin{align}
\left(  \frac{\alpha_{e}}{\alpha_{e}+\beta_{e}}\right)  ^{2} &  =\frac
{2\alpha^{2}}{\omega_{1}\omega_{2}}\label{o3hsd}\\
(1+\frac{\beta_{e}}{\alpha_{e}}\exp(-\left(  \alpha_{e}+\beta_{e}\right)
t))^{2} &  =\frac{\overline{O}_{3}(t)}{\overline{O}_{3s}}.\label{o3htd}%
\end{align}
If  $\alpha\approx0$,  from Eqs. (\ref{char1}), (\ref{o3hsd}) and
(\ref{o3htd}), $\omega_{1}\approx\beta k_{c}/(k_{o\text{ }}+k_{c}%
),$\ \ $\omega_{2}\approx\beta$, $\alpha_{e}\approx0,$ and $\beta_{e}%
\approx\beta k_{c}/(k_{o\text{ }}+k_{c})$. Therefore, the solution $O_{4}%
(t)$\ of Eqs. (\ref{open1}) to (\ref{open4})\ for K channel activation and
deactivation is proportional to \ $O_{3H}(t)$ (see Fig. 3), and the derived
rates \ $\alpha_{e}$ and \ $\beta_{e}$ \ have a voltage dependence similar to
the rates $\alpha_{n}$ and $\beta_{n}$ (see Figs. 4(a) and (b)).

\textbf{A KINETIC MODEL WITH TWO STAGE ACTIVATION OF VOLTAGE SENSORS AND AN
OPENING STEP}

The empirical forward rate $\alpha_{n}$ between the closed state and the open
state for the K channel voltage sensor is not an exponential function of the
membrane potential $V$ \cite{hh}, and therefore, it is assumed that each
voltage sensor activates in 2 stages \ with forward rates $\alpha_{s}%
(V)\ $\ and $\gamma_{s}(V)$, and backward rates$\ \beta_{s}(V)$ and
$\ \ \delta_{s}(V)$\ ($s=1,2$). Two stage voltage sensor activation is
supported by a rising phase in the K channel gating current \cite{wb} and a
kinetic model description of the ionic and gating currents of the Shaker IR K
channel \cite{zha}. \ For a K channel with 2 non-identical independent
voltage sensors, and a separate opening step following activation,\ \ the
occupation probability of the open state \ $O_{33}$, and the closed states
\ $C_{ij}$ for $i=1,3$ and $j=1,3$, \ are determined by the kinetic equation
\ (see Fig. 5)%

\begin{equation}
\frac{dC_{11}}{dt}=-(\alpha_{1}+\alpha_{2})C_{11}(t)+\beta_{1}C_{21}%
(t)+\beta_{2}C_{12}(t), \label{10c1}%
\end{equation}

\begin{equation}
\frac{dC_{21}}{dt}=-(\beta_{1}+\gamma_{1}+\alpha_{2})C_{21}(t)+\alpha
_{1}C_{11}(t)+\delta_{1}C_{31}(t)+\beta_{2}C_{22}(t), \label{10c2}%
\end{equation}

\begin{equation}
\frac{dC_{31}}{dt}=\gamma_{1}C_{21}(t)-(\delta_{1}+\alpha_{2})C_{31}%
(t)+\beta_{2}C_{32}(t), \label{10c3}%
\end{equation}

\begin{equation}
\frac{dC_{12}}{dt}=\alpha_{2}C_{11}(t)-(\beta_{2}+\gamma_{2}+\alpha_{1}%
)C_{12}(t)+\beta_{1}C_{22}(t)+\delta_{2}C_{13}(t), \label{10c4}%
\end{equation}

\begin{equation}
\frac{dC_{22}}{dt}=\alpha_{2}C_{21}(t)-(\beta_{1}+\beta_{2}+\gamma_{1}%
+\gamma_{2})C_{22}(t)+\delta_{1}C_{32}(t)+\delta_{2}C_{23}(t)+\alpha_{1}%
C_{12}(t), \label{10c5}%
\end{equation}

\begin{equation}
\frac{dC_{32}}{dt}=\alpha_{2}C_{31}(t)-(\beta_{2}+\gamma_{2}+\delta_{1}%
)C_{32}(t)+\gamma_{1}C_{22}(t)+\delta_{2}C_{33}(t), \label{10c6}%
\end{equation}

\begin{equation}
\frac{dC_{13}}{dt}=\gamma_{2}C_{12}(t)-(\alpha_{1}+\delta_{2})C_{13}%
(t)+\beta_{1}C_{23}(t), \label{10c7}%
\end{equation}

\begin{equation}
\frac{dC_{23}}{dt}=\gamma_{2}C_{22}(t)-(\beta_{1}+\gamma_{1}+\delta_{2}%
)C_{23}(t)+\delta_{1}C_{33}(t)+\alpha_{1}C_{13}(t), \label{10c8}%
\end{equation}

\begin{equation}
\frac{dC_{33}}{dt}=\gamma_{2}C_{32}(t)-(\delta_{1}+\delta_{2}+k_{o}%
)C_{33}(t)+\gamma_{1}C_{23}(t)+k_{c}O_{33}(t). \label{10c9}%
\end{equation}

\begin{equation}
\frac{dO_{33}}{dt}=k_{o}C_{33}(t)-k_{c}O_{33}(t) \label{10o10}%
\end{equation}
where $k_{o\text{ }}$ and $k_{c}$, are the opening and closing rates, and the
voltage-dependent rates are $\alpha_{s}(V)=\alpha_{s0}\exp(q_{\alpha_{s}%
}V/kT)$, $\ \ \beta_{s}(V)=\beta_{s0}\exp(q_{\beta_{s}}V/kT)$, $\ \gamma
_{s}(V)=\gamma_{s0}\exp(q_{\gamma_{s}}V/kT)$, $\delta_{s}(V)=\delta_{s0}%
\exp(q_{\delta_{s}}V/kT)$, where $q_{s}$\ is the equivalent charge for each
transition, and k is Boltzmann's constant. If the voltage sensors are
independent, \ the occupation probabilities of the closed states $C_{ij}$
$=P_{1i}P_{2j},$\ \ where $P_{1i}$ represents the \ occupation probability
\ of the first voltage sensor for $i=1,3$, and $P_{2j}$ represents the
\ occupation probability of the second voltage sensor for $j=1,3$.

If $\gamma_{s}\gg\alpha_{s}$ and $\beta_{s}\gg\delta_{s}$, $\ C_{12}$,
$C_{21}$, $C_{23}$, $C_{23}$ and $C_{32}$ rapidly attain a quasi steady state
before the other states and $C_{22}\approx0,$ and therefore%

\begin{align}
C_{12}  &  \approx\frac{\alpha_{2}C_{11}+\delta_{2}C_{13}}{\beta_{2}%
+\gamma_{2}}\label{c12}\\
C_{21}  &  \approx\frac{\alpha_{1}C_{11}+\delta_{1}C_{31}}{\beta_{1}%
+\gamma_{1}}\label{c21}\\
C_{23}  &  \approx\frac{\alpha_{1}C_{13}+\delta_{1}C_{33}}{\beta_{1}%
+\gamma_{1}}\label{c23}\\
C_{32}  &  \approx\frac{\alpha_{2}C_{31}+\delta_{2}C_{33}}{\beta_{2}%
+\gamma_{2}}. \label{c32}%
\end{align}
By substitution into \ Eqs. (\ref{10c1}) to (\ref{10o10}), \ the 10 state
kinetic equation may be reduced to the 5 state kinetic scheme (see Figs. 6,
7(a) and (b)))%

\begin{equation}
\frac{dC_{11}}{dt}=-(\alpha_{1r}+\alpha_{2r})C_{11}(t)+\beta_{1r}%
C_{31}(t)+\beta_{2r}C_{13}(t), \label{5c1}%
\end{equation}

\begin{equation}
\frac{dC_{31}}{dt}=\alpha_{1r}C_{11}(t)-(\beta_{1r}+\alpha_{2r})C_{31}%
(t)+\beta_{2r}C_{33}(t), \label{5c2}%
\end{equation}

\begin{equation}
\frac{dC_{13}}{dt}=\alpha_{2r}C_{11}(t)-(\alpha_{1r}+\beta_{2r})C_{13}%
(t)+\beta_{1r}C_{33}(t), \label{5c3}%
\end{equation}

\begin{equation}
\frac{dC_{33}}{dt}=\alpha_{2r}C_{31}(t)-(\beta_{1r}+\beta_{2r}+k_{o}%
)C_{33}(t)+\alpha_{1r}C_{13}(t)+k_{c}O_{33}(t), \label{5c4}%
\end{equation}

\begin{equation}
\frac{dO_{33}}{dt}=k_{o}C_{33}(t)-k_{c}O_{33}(t). \label{5o5}%
\end{equation}
where for each $s$, $\alpha_{sr}$ and $\beta_{sr}$ are the reduced rates%

\begin{align}
\alpha_{sr}  &  \approx\frac{\alpha_{s}\gamma_{s}}{\beta_{s}+\gamma_{s}%
}\label{reduce1}\\
\beta_{sr}  &  \approx\frac{\beta_{s}\delta_{s}}{\beta_{s}+\gamma_{s}}.
\label{reduce2}%
\end{align}

If each voltage sensor is identical and has transition rates $\alpha
_{s}=\alpha$, $\beta_{s}=\beta$, $\gamma_{s}=\gamma$, $\delta_{s}=\delta$, for
$s=1,2$, \ and $k_{o\text{ }}\gg$\ $\alpha,\gamma$ $\ $and $k_{c\text{ }}$
$\gg$\ $\beta,\delta$, \ \ Eqs. (\ref{10c1}) to (\ref{10o10}) \ may be reduced
to
\begin{equation}
\frac{dC_{11}}{dt}=-2\alpha C_{11}(t)+\beta(C_{21}(t)+C_{12}(t)), \label{9c1}%
\end{equation}

\begin{equation}
\frac{dC_{21}}{dt}=-(\gamma+\beta+\alpha)C_{21}(t)+\alpha C_{11}(t)+\delta
C_{31}(t)+\beta C_{22}(t), \label{c2}%
\end{equation}

\begin{equation}
\frac{dC_{31}}{dt}=\gamma C_{21}(t)-(\delta+\alpha)C_{31}(t)+\beta C_{32}(t),
\label{9c3}%
\end{equation}

\begin{equation}
\frac{dC_{12}}{dt}=\alpha C_{11}(t)-(\beta+\gamma+\alpha)C_{12}(t)+\beta
C_{22}(t)+\delta C_{13}(t), \label{9c4}%
\end{equation}

\begin{equation}
\frac{dC_{22}}{dt}=\alpha(C_{21}(t)+C_{12}(t))-2(\beta+\gamma)C_{22}%
(t)+\delta(C_{32}(t)+C_{23}(t)), \label{9c5}%
\end{equation}

\begin{equation}
\frac{dC_{32}}{dt}=\alpha C_{31}(t)-(\beta+\gamma+\delta)C_{32}(t)+\gamma
C_{22}(t)+\frac{\delta k_{c}}{k_{o\text{ }}+k_{c}}\overline{O}_{33}(t),
\label{9c6}%
\end{equation}

\begin{equation}
\frac{dC_{13}}{dt}=\gamma C_{12}(t)-(\alpha+\delta)C_{13}(t)+\beta C_{23}(t),
\label{9c7}%
\end{equation}

\begin{equation}
\frac{dC_{23}}{dt}=\gamma C_{22}(t)-(\gamma+\beta+\delta)C_{23}(t)+\frac
{\delta k_{c}}{k_{o\text{ }}+k_{c}}\overline{O}_{33}(t)+\alpha C_{13}(t),
\label{9c8}%
\end{equation}

\begin{equation}
\frac{d\overline{O}_{33}}{dt}=\gamma(C_{32}(t)+C_{23}(t))-\frac{2\delta k_{c}%
}{k_{o\text{ }}+k_{c}}\overline{O}_{33}(t), \label{9o9}%
\end{equation}
and Eqs. (\ref{5c1}) to (\ref{5o5}) \ may be reduced to%

\begin{equation}
\frac{dC_{11}}{dt}=-2\alpha_{r}C_{11}(t)+\beta_{r}(C_{31}(t)+\beta C_{13}(t)),
\label{4c1}%
\end{equation}

\begin{equation}
\frac{dC_{31}}{dt}=\alpha_{r}C_{11}(t)-(\alpha_{r}+\beta_{r})C_{31}%
(t)+\frac{k_{c}\beta_{r}}{k_{o\text{ }}+k_{c}}\overline{O}_{33}(t),
\label{4c2}%
\end{equation}

\begin{equation}
\frac{dC_{13}}{dt}=\alpha_{r}C_{11}(t)-(\alpha_{r}+\beta_{r})C_{13}%
(t)+\frac{k_{c}\beta_{r}}{k_{o\text{ }}+k_{c}}\overline{O}_{33}(t),
\label{4c3}%
\end{equation}

\begin{equation}
\frac{d\overline{O}_{33}}{dt}=\alpha_{r}(C_{31}(t)+C_{13}(t))-\frac
{2k_{c}\beta_{r}}{k_{o\text{ }}+k_{c}}\overline{O}_{33}(t),\label{4o4}%
\end{equation}
where $\overline{O}_{33}(t)=C_{33}+O_{33}$, and from the characteristic
equation\  for each  voltage sensor,\ \ the reduced rates $\alpha_{r}$ and
$\beta_{r}$ \ are%

\begin{align*}
\alpha_{r} &  =\frac{\alpha\gamma}{\omega_{2}}\approx\frac{\alpha\gamma}%
{\beta+\gamma}\\
\beta_{r} &  =\frac{\delta(\alpha+\beta)}{\omega_{2}}\approx\frac
{\delta(\alpha+\beta)}{\beta+\gamma}\\
\omega_{1} &  =\alpha_{r}+\beta_{r}\\
\omega_{2} &  \approx\beta+\gamma+\frac{\alpha\beta+\delta(\gamma-\alpha
)}{\beta+\gamma}\approx\beta+\gamma.
\end{align*}

By defining $\overline{C}_{13}(t)=C_{13}(t)+C_{31}(t)$, Eqs. (\ref{4c1}) to
(\ref{4o4}) \ may be expressed as%

\begin{equation}
\frac{dC_{11}}{dt}=-2\alpha_{r}C_{11}(t)+\beta_{r}\overline{C}_{13}(t),
\label{4s1}%
\end{equation}

\begin{equation}
\frac{d\overline{C}_{13}}{dt}=2\alpha_{r}C_{11}(t)+\frac{2k_{c}\beta_{r}%
}{k_{o\text{ }}+k_{c}}\overline{O}_{33}(t)-(\alpha_{r}+\beta_{r})\overline
{C}_{13}(t), \label{4s2}%
\end{equation}

\begin{equation}
\frac{d\overline{O}_{33}}{dt}=\alpha_{r}\overline{C}_{13}(t)-\frac{2k_{c}%
\beta_{r}}{k_{o\text{ }}+k_{c}}\overline{O}_{33}(t),\label{4s3}%
\end{equation}
As in the previous Section, for activation of a K channel during a voltage
clamp, $\overline{O}_{33}(t)$ may be approximated by \ the solution
$O_{33H}(t)$\ of the kinetic equation%

\begin{equation}
\frac{dC_{11}}{dt}=-2\alpha_{e}C_{11}(t)+\beta_{e}\overline{C}_{13}(t),
\label{3 state 2 stage hh1}%
\end{equation}

\begin{equation}
\frac{d\overline{C}_{13}}{dt}=2\alpha_{e}C_{11}(t)+2\beta_{e}O_{33H}(t)
-(\alpha_{e}+\beta_{e})\overline{C}_{13}(t), \label{3 state 2 stage hh2}%
\end{equation}

\begin{equation}
\frac{dO_{33H}}{dt}=\alpha_{e}\overline{C}_{13}(t)-2\beta_{e}O_{33H}%
(t),\label{3 state 2 stage hh3}%
\end{equation}
where the effective rate functions \ $\alpha_{e}>\beta_{e}$ may be determined
from equations similar to Eqs. (\ref{o3hs}) and (\ref{o3ht}) for activation,
and if $\beta\approx0$, \ %

\[
\alpha_{e}\approx\alpha_{r}\approx\frac{\alpha\gamma}{\beta+\gamma}.
\]
The  rates \ $\alpha_{e}< \beta_{e}$ may be determined from  Eqs.
(\ref{o3hsd}) and (\ref{o3htd}) for deactivation, and if $\alpha\approx0$, \ %

\begin{equation}
\beta_{e}\approx\frac{k_{c}}{k_{o\text{ }}+k_{c}}\beta_{r}=\frac{k_{c}%
}{k_{o\text{ }}+k_{c}}\frac{\delta(\alpha+\beta)}{\beta+\gamma}.\label{eff2}%
\end{equation}
The forward rate \ $\alpha_{e}$\ has an exponential quasilinear voltage
dependence if $\ $the\ equivalent ch$\arg$e $\ q_{\alpha}\approx0.3e$, and
\ $\beta_{e}$ is approximately exponential but the voltage sensor rate is
reduced by the factor \ $k_{c}/(k_{o\text{ }}+k_{c})$, \ \ and can account for
the voltage dependence of $\alpha_{n}$ and $\beta_{n}$ (see Fig. 8).
Therefore, during activation the open state probability $O_{33}(t)$ \ of a 10
state kinetic model of a K channel with 2 stage activation for each voltage
sensor and an opening step may be approximated by $\ (k_{o\text{ }%
}/(k_{c\text{ }}+k_{o\text{ }}))O_{33H}(t)\ $ (see Figs. 9(a) and (b)).

\bigskip

\pagebreak

\textbf{CONCLUSION}

HH described the time-dependence of the K current \ during a voltage clamp for
activation (when $\alpha_{n}>\beta_{n}$ ) and for deactivation (when
$\alpha_{n}<\beta_{n}$ ) \ by the expression $n^{4}$ where \ $n$ $\ $satisfies
a rate quation.$\ $\ The empirical description of the K current, as well as
the Na current, was sufficiently accurate to calculate a membrane action
potential as well as a threshold potential and refractory period, and there
was  reasonable agreement with experimental data \cite{hh}. However, the HH
model does not account for the rising phase in the K channel gating current,
the non-exponential voltage dependence of the rate function $\alpha_{n}$ \ and
the asymmetry of the\ voltage dependence of \ the rate functions $\alpha_{n}$
and $\beta_{n}$.

In this paper, it has been shown that a kinetic model of a K channel with two
voltage sensors, and a separate opening step,\ has 4 states and may be reduced
to a 3 state system where the effective voltage sensor rate function from the
activated-open state \ to the first closed state is $2\beta k_{o\text{ }%
}/(k_{c}+k_{o\text{ }})$ where $k_{o\text{ }}$ $\gg k_{c}$ are the opening and
closing rates, and \ $\beta$\ \ is the voltage sensor rate. If each voltage
sensor is activated in two stages with forward rates $\gamma\gg\alpha\ $, and
backward rates $\ \beta\gg\delta$, and there is a separate opening step, \ the
occupation probabilities of the states of the channel satisfy a 10 state
kinetic equation that may be reduced to a 3 state model with activated-open
state probability $\overline{O}_{3}(t)$.\ \ If the effective rates $\alpha
_{e}\ $and $\beta_{e}$ of the rate equation in \ $n(t)$\ satisfy $\overline
{O}_{3}(t)=$ $n(t)^{2}$ \ for activation and deactivation, $\alpha_{e}\ $and
$\beta_{e}$ have a similar voltage dependence to $\alpha_{n}$ and $\beta_{n}$
from voltage clamp data. 

A kinetic model with two stage activation of\ each of the four voltage
sensors, and a separate opening step, may be reduced to a five state model
with activated open probability $\overline{O}_{5}(t)$. Assuming that the
kinetic equation that describes the activation of four independent two-state
voltage sensors has open probability $O_{5H}(t)$, the HH model is obtained as
an approximation from the determination of rate functions $\alpha_{e}$ and
$\beta_{e}$ that satisfy the equation $\overline{O}_{5}(t)=O_{5H}(t)$ for
activation when $\alpha_{e}$ $>$\ $\beta_{e}$\ \ and for deactivation when
$\alpha_{e}$ $\leq$\ $\beta_{e}$.

\pagebreak

\begin{figure*}
\begin{center}
\includegraphics[width=0.7\textwidth]{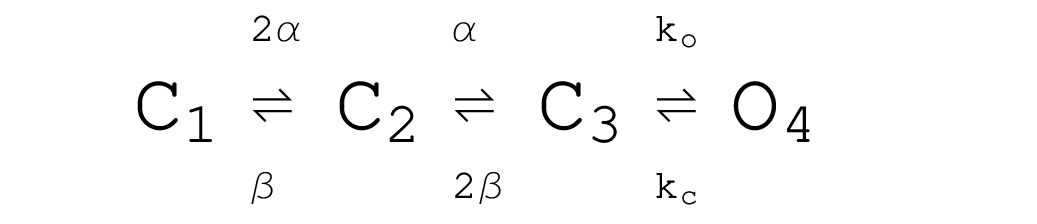}
\caption{State diagram for activation of a K channel with 2 identical independent
voltage sensors and an opening step where $\alpha$ and $\beta$ are the voltage
dependent rates between closed and activated states of the voltage sensor, and
$k_{o}$ and $k_{c}$ are the opening and closing rates between the closed and
open states with occupational probabilities $C_{3}(t)$ and $O_{4}(t)$.}
\end{center}
\end{figure*}

\begin{figure*}
\begin{center}
\includegraphics[width=0.7\textwidth]{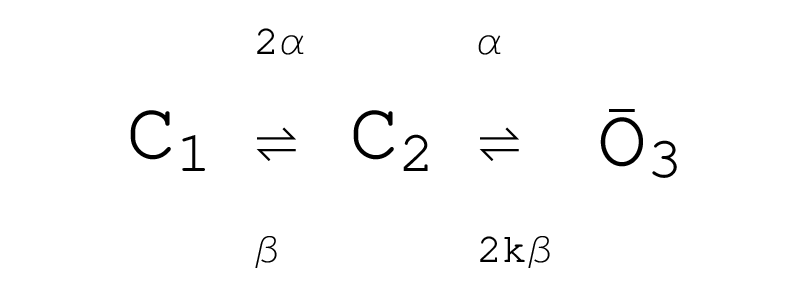}
\caption{The state diagram for activation of a K channel in Fig. 1 may be reduced to a
three state model when the opening and closing rates $\ k_{o}$ and $k_{c}$
satisfy $k_{o\text{ }}$ $\gg k_{c}$,\ $k_{o\text{ }}\gg\alpha$, $k_{c\text{ }%
}\gg\beta$, \ where \ $k=$ $k_{c\text{ }}/(k_{c\text{ }}+k_{o\text{ }}%
)$,\ \ \ $\overline{O}_{3}=C_{3\text{ }}+O_{4}$\ \ and the occupational
probabilities $C_{3\text{ }}$ and $O_{4}$ \ rapidly attain a quasi-steady state.}
\end{center}
\end{figure*}

\begin{figure*}
\begin{center}
\includegraphics[width=0.7\textwidth]{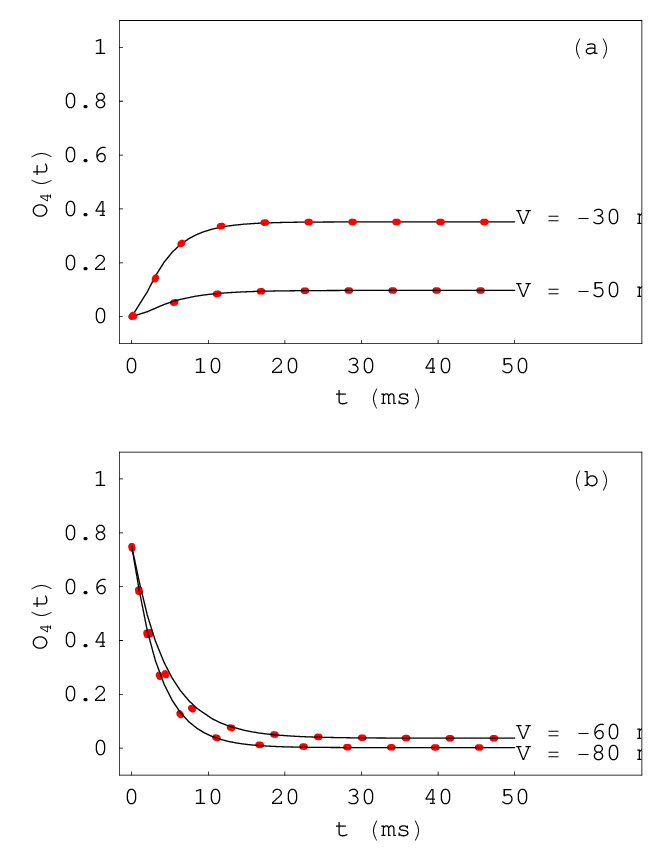}
\caption{During the voltage clamp of a K channel, (a) the occupational probability
$O_{4}(t)$ from Eqs. (\ref{open1}) to \ (\ref{open4}) during activation (solid
line) is approximated by $O_{3H}(t)=\left(  \alpha_{e}/(\alpha_{e}+\beta
_{e})\right)  ^{2}(1-\exp(-\omega_{e}t))^{2}$ \ (dotted line) \ and (b) the
occupational probability $O_{4}(t)$ during deactivation (solid line) is
approximated by $O_{3H}(t)=(\alpha_{e}+\beta_{e}\exp(-\left(  \alpha_{e}%
+\beta_{e}\right)  t))^{2}/\left(  \alpha_{e}+\beta_{e}\right)  ^{2}$
\ (dotted line) where the voltage sensor rate functions $(ms^{-1})$ are
\ $\alpha=1.1\alpha_{n}$, $\beta=4\beta_{n}$ and $\alpha_{n}%
=0.01(V+50)/(1-\exp(-0.1(V+50)))$, \ $\beta_{n}=1.25\exp(-(V+60)/25)$,
$k_{o}=6$ and $k_{c}=2.$}
\end{center}
\end{figure*}

\begin{figure*}
\begin{center}
\includegraphics[width=0.7\textwidth]{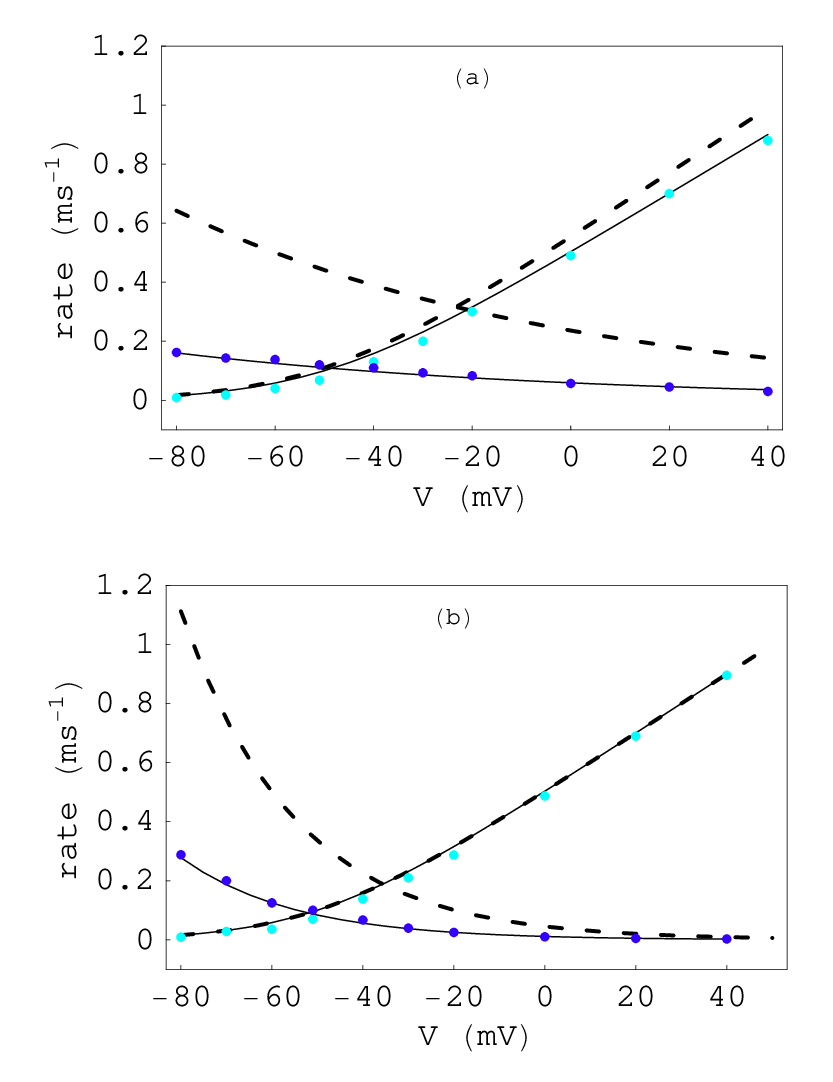}
\caption{The effective rate functions $\alpha_{e}$ and $\beta_{e}$ \ (dotted line)
\ \ are determined from Eqs. (\ref{o3hs}) and (\ref{o3ht}) for membrane
depolarisation when $\alpha_{e}>\beta_{e}$, and from Eqs. (\ref{o3hsd}) and
(\ref{o3htd}) for membrane hyperpolarisation when $\alpha_{e}<\beta_{e}$, and
provide a good fit to the HH rate functions \ (solid line) \ \ where the
voltage sensor rate functions for the kinetic model of Fig. 1 are (a)
$\alpha=1.1\alpha_{n}$, $\beta=4\beta_{n}$ (dashed line) \ where $\alpha
_{n}=0.01(V+50)/(1-\exp(-0.1(V+50)))$ and $\beta_{n}=1.25\exp(-(V+60)/80)$ and
(b) $\alpha=\alpha_{n}$, $\beta=4\beta_{n}$ (dashed line) \ where $\alpha
_{n}=0.01(V+50)/(1-\exp(-0.1(V+50)))$ and $\ \beta_{n}=1.25\exp(-(V+60)/25)$,
$k_{o}=6$ and $k_{c}=2$.}
\end{center}
\end{figure*}

\begin{figure*}
\begin{center}
\includegraphics[width=0.7\textwidth]{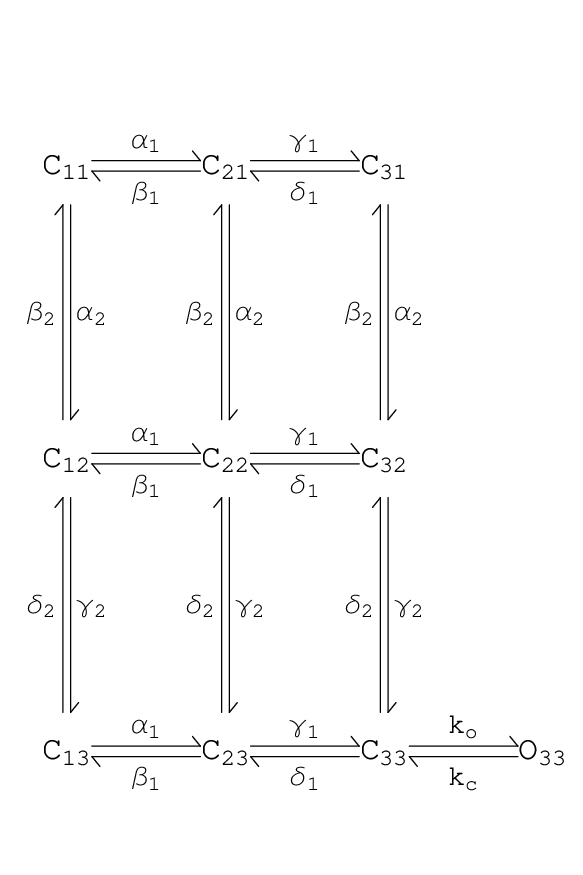}
\caption{State diagram for two stage activation of a K channel where horizontal
transitions represent the activation of the first voltage sensor and vertical
transitions represent the activation of the second voltage sensor, and the
opening step is from the activated state to the open state with occupational
probabilities $C_{33\text{ }}$ and $O_{33}$.}
\end{center}
\end{figure*}

\begin{figure*}
\begin{center}
\includegraphics[width=0.7\textwidth]{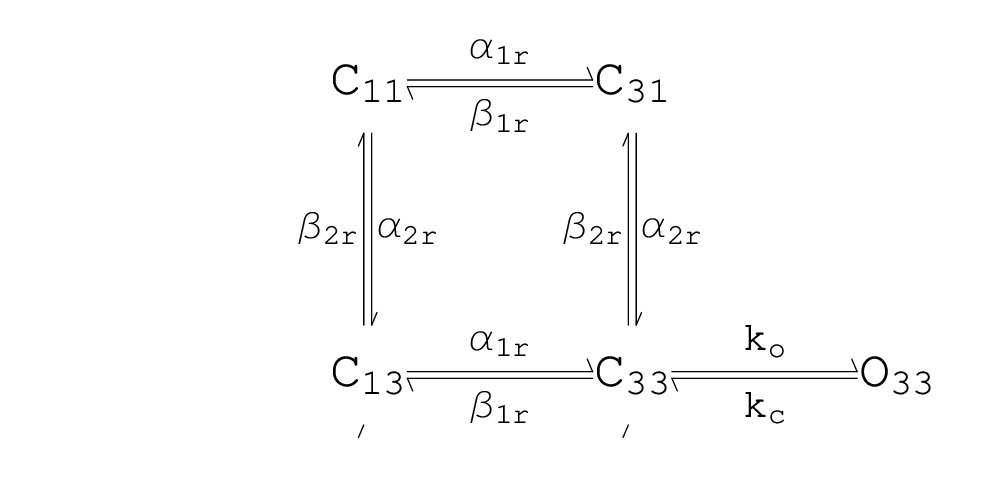}
\caption{The ten state kinetic model for K channel activation in Fig. 5 may be reduced
to a five state model when $\gamma_{s}\gg\alpha_{s}$ and $\beta_{s}\gg
\delta_{s}$, where the derived rate functions \ $\alpha_{sr}$ and $\beta_{sr}$
are given by Eqs. (\ref{reduce1}) to (\ref{reduce2}).}
\end{center}
\end{figure*}

\begin{figure*}
\begin{center}
\includegraphics[width=0.7\textwidth]{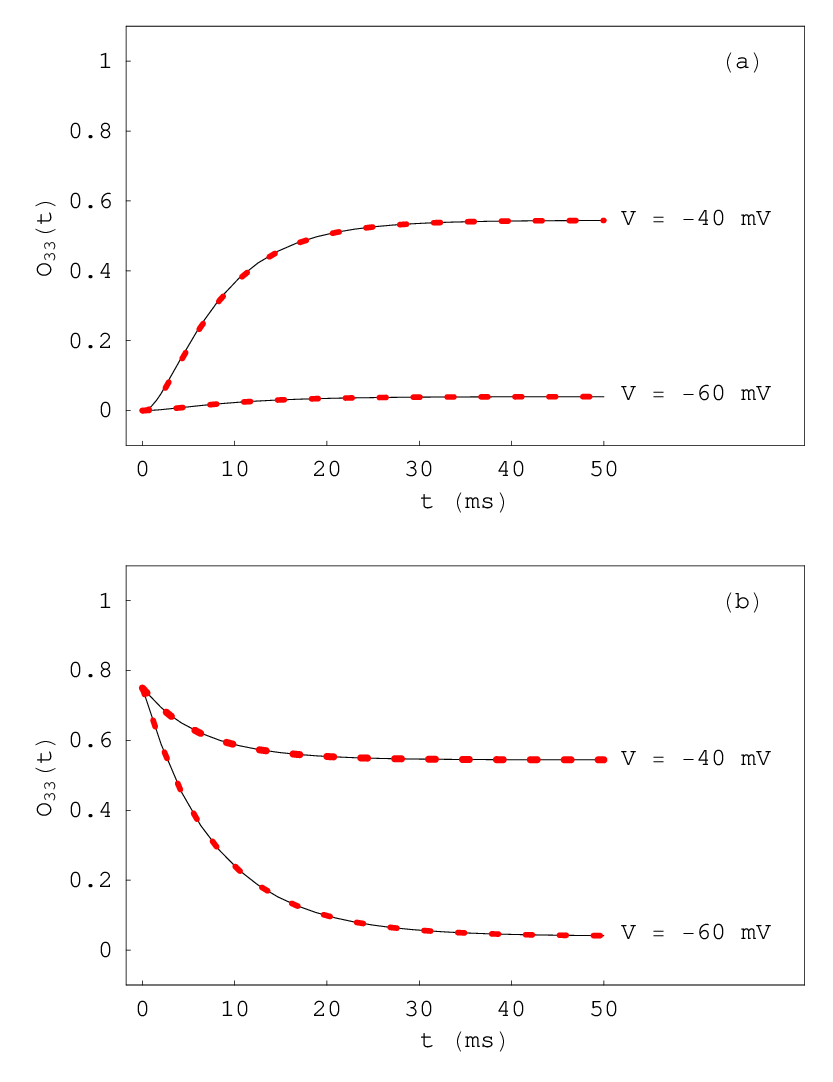}
\caption{During the voltage clamp of a K channel, the open probability $O_{33}(t)$
\ from a ten state model \ of K channel gating (solid line) and\ $O_{33}(t)$
\ \ from the reduced 5 state model (dotted line) during (a) activation and (b)
deactivation where the rate functions for the K channel $(ms^{-1})$
are\ \ $\alpha=1.1\exp(0.3V/25)$, $\gamma=2.8\exp(0.32V/25)$, $\beta
=0.37\exp(-1.6V/25)$, $\delta=0.02\exp(-1.1V/25)$, $k_{o}=6$ and $k_{c}=2.$}
\end{center}
\end{figure*}

\begin{figure*}
\begin{center}
\includegraphics[width=0.7\textwidth]{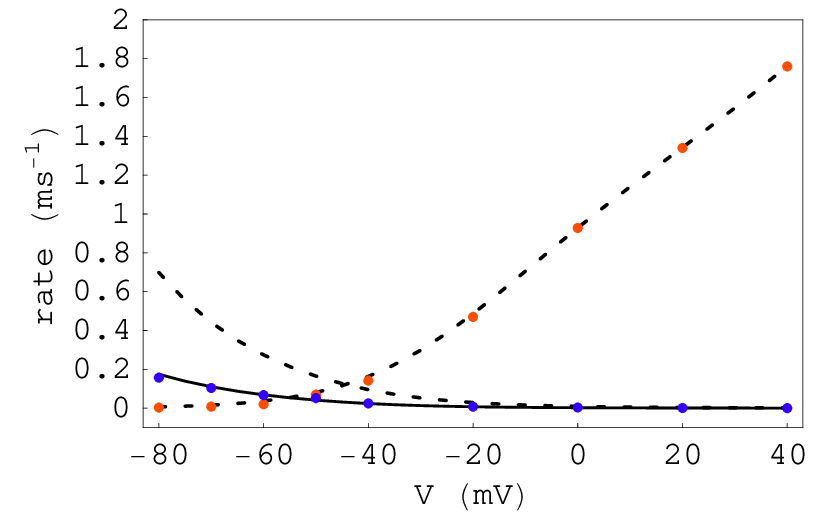}
\caption{For a K channel with 2 voltage sensors, 
two stage activation of each voltage sensor, and an opening step,
the derived rate functions $\alpha_{r}$ and $\beta_{r}$  from Eqs. (\ref{reduce1}) to
 (\ref{reduce2}) \ (dashed line), the rate function
$\ \beta_{r}k_{o\text{ }}/(k_{c\text{ }}+k_{o\text{ }})$ \ \ (solid line) and
the effective rate functions $\alpha_{e}$ and $\beta_{e}$ determined from Eqs
(\ref{o3hs}) to (\ref{o3ht}) \ for activation, \ and Eqs. (\ref{o3hsd}) and
(\ref{o3htd}) for deactivation (dotted line). The rate functions for the K
channel $(ms^{-1})$ are$\ \ \alpha=1.1\exp(0.3V/25)$, $\ \gamma=2.8\exp
(0.32V/25)$, $\ \beta=0.37\exp(-1.6V/25)$, $\ \delta=0.02\exp(-1.1V/25)$,
$k_{o}=6$ and $k_{c}=2.$}
\end{center}
\end{figure*}

\begin{figure*}
\begin{center}
\includegraphics[width=0.7\textwidth]{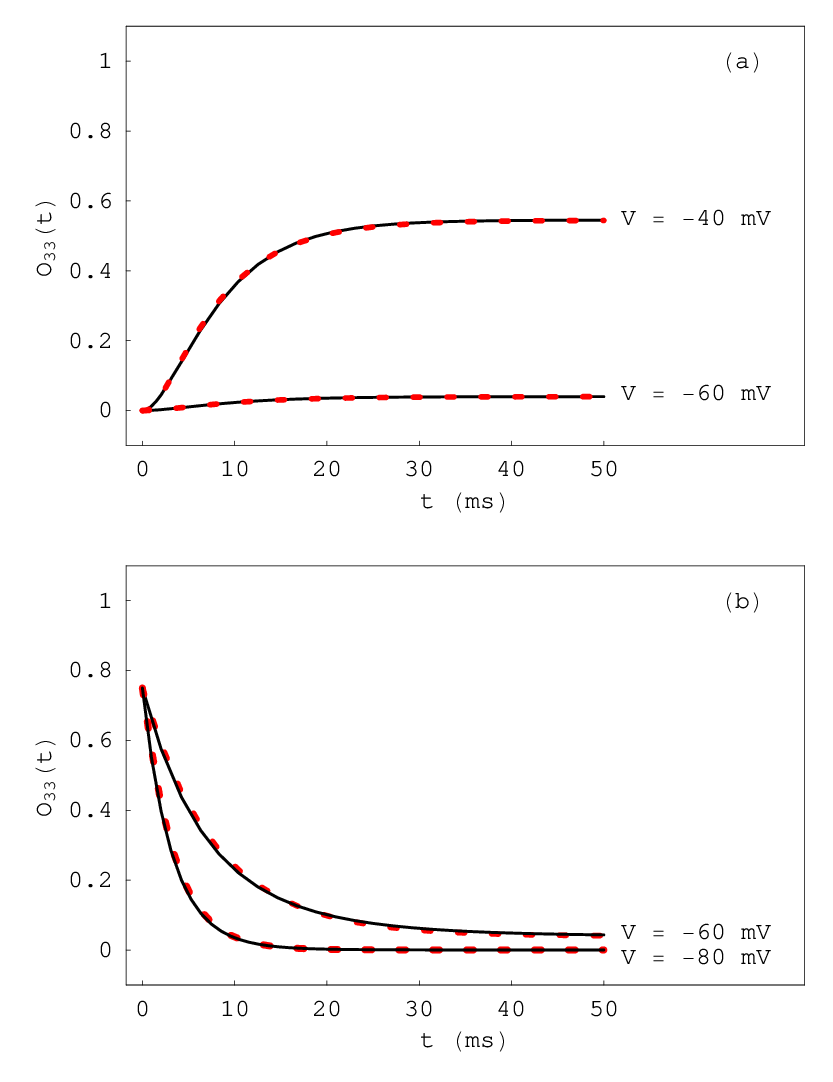}
\caption{During the voltage clamp of a K channel, \ the open probability $O_{33}(t)$
\ from a ten state model \ of K channel gating with 2 stage activation for
each voltage sensor and an opening step (solid line) \ and \ ($k_{o\text{ }%
}/(k_{c\text{ }}+k_{o\text{ }}))O_{33H}(t)$ \ \ (dotted line) \ where
$O_{33H}(t)$\ is the solution of Eqs. (\ref{3 state 2 stage hh1}) to
(\ref{3 state 2 stage hh3}) during activation (a) and deactivation (b),
\ where the rate functions for the K channel $(ms^{-1})$ \ \ \ are
$\alpha=1.1\exp(0.3V/25)$, $\gamma=2.8\exp(0.32V/25)$, $\beta=0.37\exp
(-1.6V/25)$, $\delta=0.02\exp(-1.1V/25)$, $k_{o}=6$ and $k_{c}=2.$}
\end{center}
\end{figure*}
\end{document}